\documentclass[fleqn,usenatbib]{mnras}

\usepackage{newtxtext,newtxmath}

\usepackage[T1]{fontenc}

\DeclareRobustCommand{\VAN}[3]{#2}
\let\VANthebibliography\thebibliography
\def\thebibliography{\DeclareRobustCommand{\VAN}[3]{##3}\VANthebibliography}

\usepackage{graphicx}	
\usepackage{amsmath}	
\usepackage{longtable}  
\usepackage{multirow}
\usepackage{manyfoot}
\usepackage{lipsum}
\usepackage{mathtools}
\usepackage{mathrsfs}
\usepackage{placeins}
\usepackage{booktabs}
\usepackage{xcolor}
\usepackage{lscape}
\usepackage{rotating}
\usepackage{multirow}
\usepackage{graphicx}
\usepackage{threeparttable}
\usepackage{makecell} 

\newcommand{\msol}{M$_\odot$}
\definecolor{mygreen}{rgb}{0,0.502,0}
\definecolor{myblue}{rgb}{0.07058824, 0.07058824, 0.5372549}
\hypersetup{linkcolor=mygreen,citecolor=mygreen,urlcolor=mygreen}\newcommand{\name}{4U 1636$-$536}

\title[PPM of superburst oscillations]{Impact of Accretion Assumptions on Pulse Profile Modelling of Superburst Oscillations in 4U 1636-536}

\author[Kini~et~al.]{Yves~Kini$^{1}$
\thanks{E-mail: \href{mailto:y.kini@uva.nl}{y.kini@uva.nl}},
Anna~L.~Watts$^{1}$,
Tuomo~Salmi$^{1,2}$, 
Anna Bilous$^{3}$,
Serena~Vinciguerra$^{1}$, \newauthor 
Sebastien~Guillot$^{4,5}$, 
David R. Ballantyne$^{6}$, 
Erik Kuulkers$^{7}$,
Slavko~Bogdanov$^{8}$, \newauthor 
Valery~Suleimanov$^{10}$
\\
$^{1}$Anton Pannekoek Institute for Astronomy, University of Amsterdam, Science Park 904, 1098XH Amsterdam, the Netherlands\\
$^{2}$Department of Physics, P.O. Box 64, FI-00014 University of Helsinki, Finland\\
$^{3}$Independent researcher\\
$^{4}$IRAP, CNRS, 9 avenue du Colonel Roche, BP 44346, F-31028 Toulouse Cedex 4, France \\
$^{5}$Universit\'{e} de Toulouse, CNES, UPS-OMP, F-31028 Toulouse, France\\
$^{6}$Center for Relativistic Astrophysics, School of Physics, Georgia Institute of Technology, 837 State Street, Atlanta, GA 30332-0430, USA\\
$^{7}$European Space Agency, ESTEC, Keplerlaan 1, NL-2200 AG, Noordwijk, The Netherlands.\\
$^{8}$Columbia Astrophysics Laboratory, Columbia University, 550 West 120th Street, New York, NY 10027, USA\\
$^{10}$Institut f\"ur Astronomie und Astrophysik, Kepler Center for Astro and Particle Physics, Universit\"at T\"ubingen, Sand 1, D-72076 T\"ubingen, Germany
}

\date{Accepted XXX. Received YYY; in original form ZZZ}

\pubyear{20XX}

\begin{document}
\label{firstpage}
\pagerange{\pageref{firstpage}--\pageref{lastpage}}
\maketitle

\begin{abstract}

Modelling the coherent pulsations observed during thermonuclear bursts offers a valuable method to probe the poorly understood equation of state of dense and cold matter. Here we apply the pulse profile modelling technique to the pulsations observed with RXTE during the 2001 superburst of \name. By employing a single, uniform-temperature hot spot model with varying size and temperature, along with various assumptions for background/accretion contribution, we find that each assumption leads to different inferred mass, radius, and compactness constraints. This highlights the critical need to better understand the mass accretion rate enhancement/reduction during thermonuclear bursts to accurately model burst oscillation sources using pulse profile modelling.

\end{abstract}


\section{Introduction}\label{sec:intro}

Type I X-ray bursts are thermonuclear explosions on neutron stars (NSs) in low-mass X-ray binaries \citep[LMXBs; see][for a recent review]{Galloway:2020}.  These bursts result from the accumulation of hydrogen or helium from the companion star, which forms an ocean on the surface of the NS. As pressure and temperature build-up at the base of the ocean, the conditions eventually reach a critical threshold, triggering runaway thermonuclear reactions \citep[for reviews see][]{Bildsten:1998, Keek:2016} which produce sudden and intense bursts of X-rays.

 Stable hydrogen or helium burning as well as these thermonuclear bursts can leave behind a layer of carbon-rich ashes \citep{Keek:2016}. Over time, when the pressure and temperature at the base of this carbon layer become sufficiently high, another runaway thermonuclear reaction of carbon fusion can occur, leading to what are known as superbursts. Superbursts are bright X-ray events that typically last several hours \cite[see][for a review]{Zand:2017} and release approximately  $10^{42}$ ergs \citep[see, e.g.,][]{Cumming:2001, Strohmayer:2002sb}. These events have now been reported in 16 neutron stars \citep{Alizai:2023} and can be used to probe system properties.

Similar to the coherent pulsations (known as burst oscillations) observed in the light curves of Type I X-ray bursts \citep[see][for a review]{Watts:2012, Bilous:2019bo}, coherent pulsations have been observed in one superburst, from \name, during its cooling phase \citep[][hereafter SM02]{Strohmayer:2002}. These coherent pulsations are powerful tools for inferring the masses and radii of thermonuclear burst oscillation (TBO) sources \citep{Bhattacharyya:2004pp,Lo:2013ava, Kini:2023, Kini:2024a, Kini:2024b} through the pulse profile modelling (PPM) technique.

PPM leverages both general and special relativistic effects to extract the NS properties encoded in the energy distribution and arrival times (rotational phase) of the observed photons. This technique has been employed to set constraints on the properties of rotation-powered millisecond pulsars \citep[see, e.g.,][]{Miller:2019,Riley:2019, Choudhury:2024,Dittmann:2024,Salmi:2024,Vinciguerra:2024} using Neutron Star Interior Composition Explorer \citep[NICER;][]{NICER} data. These constraints have not only provided unprecedented insights into the equation of state of cold and dense matter \citep{Raaijmakers:2021,Annala:2023,Takatsy:2023,Zhu:2023,Huang:2024,Pang:2024,Kurkela:2024,Rutherford:2024}, but have also shed light on the surface properties of these stars' magnetic field structure \citep{Bilous:2019, Chen:2020, Kalapotharakos:2021, Petri:2023}.

Recent works \citep{Kini:2023,Kini:2024a} have explored methods to accurately infer the properties of TBO sources using PPM of the  pulsations observed during Type I X-ray bursts, employing the X-ray Pulse Simulation and Inference \citep[\texttt{X-PSI\footnote{https://github.com/xpsi-group/xpsi.git}};][]{Riley2023} package. By dividing bursts into short-time segments and jointly modelling these segments, the systematic bias introduced by overlooking the short-scale time variability associated with these bursts and burst oscillations can be mitigated.
A follow-up study \citep{Kini:2024b} applied these techniques to the burst oscillation source XTE~J1814$-$338, yielding a mass of $1.21^{+0.05}_{-0.05}$ \msol ~
and an equatorial radius of $7.0^{+0.4}_{-0.4}$ km. The small inferred radius, coupled with the challenge of explaining the harmonic content in the data led the authors to suggest potential limitations \citep[see Sec 4.5 of][]{Kini:2024b} with using the applied model(s). Nevertheless, this study was a useful step in the use of PPM for TBO sources, showing its potential for understanding the burst oscillation mechanism. 

The properties of NSs inferred through PPM can be sensitive to the background limits imposed during sampling \citep{Lo:2013ava, Psaltis:2014,Salmi:2022}.  In the context of PPM of TBO sources, "background" refers to counts not originating directly from the stellar surface. This includes counts not caused by any astrophysical X-ray source, as well as counts from other X-ray sources within the field of view or within the LMXB system itself.  Given that we do not yet model the accretion disk and (if present) any accretion column effect for TBO sources, photons originating from the accretion disk constitute the dominant background source. This presents a significant challenge in modelling TBO sources, as the contribution of accretion to the overall burst flux remains poorly constrained.

One approach to modelling burst spectra assumes that the accretion contribution during the burst has the same spectrum as the persistent emission, enhanced or reduced by a factor \citep[\texttt{fa}; see][]{Worpel:2013fka}. The exact upper and lower limits on the scaling factor \texttt{fa} are hard to constrain a priori, as they depend on the sources and the burst. From burst spectral modelling, values up to \texttt{fa} $\approx 80$ have been obtained, and a reduction rather than an enhancement can also be expected \citep[\texttt{fa} $\le 1$;][]{Worpel:2015, Degenaar:2018}. On the theoretical side, hydrodynamic simulations of the accretion disk's response to bursts have shown that mass accretion rates can increase by up to a factor of 10 in some cases \citep{Fragile:2018, Fragile:2020, Speicher:2023, Speicher:2024}. This enhancement is due to Poynting-Robertson drag, along with disk collapse driven by Compton cooling in the case of hot and geometrically thick accretion flows. However, \citet{Speicher:2024} found that \texttt{fa} correlates more closely with the thermal temperature of the disk surface heated by the burst, rather than the mass accretion rate itself. This makes accurately modelling the accretion contribution during bursts, particularly challenging. Further challenges may arise from the possibility that \texttt{fa} is spectrally dependent \citep{Kajava:2017}.

In the present work, we apply PPM to the pulsations observed with the Rossi X-Ray Timing Explorer Proportional Counter Arrays \citep[RXTE/PCA;][]{Jahoda:1996} during the 2001 superburst of \name \ to infer its properties. We follow the \texttt{fa} method to set limits on the accretion contribution during the superburst. Given that the scaling factor \texttt{fa} is difficult to determine a priori, we set an upper limit on \texttt{fa}, assuming various scenarios.

The rest of the paper is structured as follows. First, we discuss the modelling methodology, including the data and models used in Section \ref{sec:method}. We present our results in Section  \ref{sec:result}. In Section \ref{sec:discussion}, we explore interesting aspects of the results. Finally, we conclude in Section \ref{sec:conclusion}.

\section{Method}\label{sec:method}  

The data and reduction methods for this analysis are outlined in Section \ref{sec:data}. The model is detailed in Section \ref{sec:model} and includes the space-time framework in which the star is embedded, the surface emission, the background model, the atmosphere model, the interstellar medium (ISM) absorption model, and the instruments model.

\subsection{Data}\label{sec:data}

\begin{figure}
    \centering
    \includegraphics[width=\columnwidth]{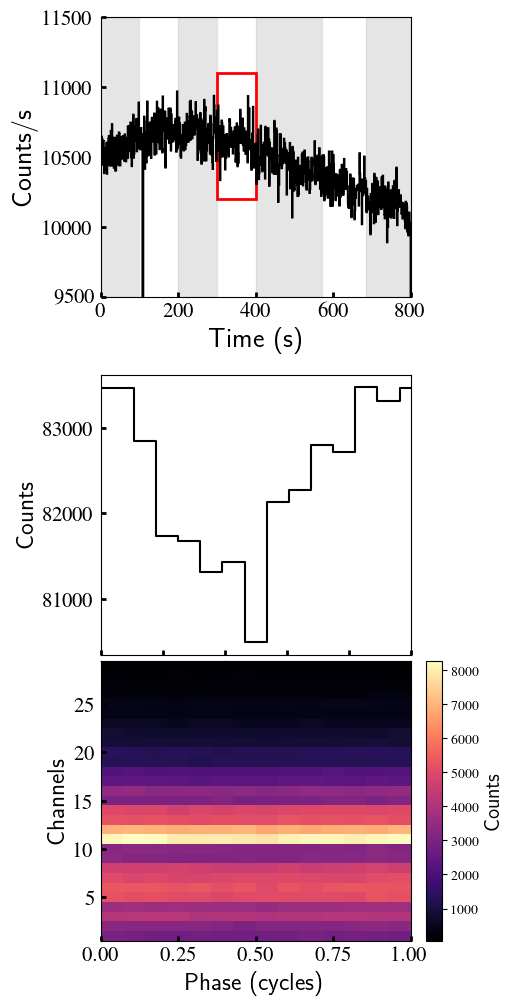}
      \caption{The portion of the light curve of \name \ showing the burst oscillation (top plot). We note that the interval where the count rate drops below the plot range falls outside a good-time interval of RXTE, and its contribution is negligible. The grey and white vertical bands indicate the data slicing used for the analysis (see Section. \ref{sec:model}).  The middle and bottom panels, respectively, show the pulse profile summed over the channel subset [1, 30] and the phase-folded pulse profile for segment 4, which is highlighted in red (top plot). The root mean square fractional amplitude (rms FA) of the pulse is approximately 0.7\% which is very low.
      }
    \label{fig:light_cures}
\end{figure}

\name \ is a persistent accreting neutron star known for its prolific Type I X-ray bursting activity, with more than 600 normal Type I X-ray bursts \citep{Galloway:2020} along with four superbursts reported \citep{Wijnands:2001, Kuulkers:2004, Kuulkers:2009, Strohmayer:2002}. In this analysis, we focus on data from the superburst observed with RXTE/PCA on 2001 February 22,  where pulsations were detected as reported by \citet{Strohmayer:2002}. The light curve of this portion of the superburst is shown in the top panel of Figure \ref{fig:light_cures}.
The data collection for this superburst began at MJD 51962 (2001 February 22) and spans approximately 800 seconds. The observations were recorded in an event mode (E\_125us\_64M\_0\_1s) with 64 energy channels and a time resolution of 122\,$\mu$s. The data were barycentered using the DE200 solar system ephemeris with the \texttt{faxbary} tool from the HEASARC package\footnote{\url{https://heasarc.gsfc.nasa.gov/}}. The source coordinates were taken to be $\mathrm{RA} = 16^\mathrm{h}40^\mathrm{m}55^\mathrm{s}.44$ and $\mathrm{DEC} = -53^\circ45^\prime05^{\prime\prime}.04$ \citep{Galloway:2020}.  

\begin{figure}
    \centering
    \includegraphics[width=\columnwidth]{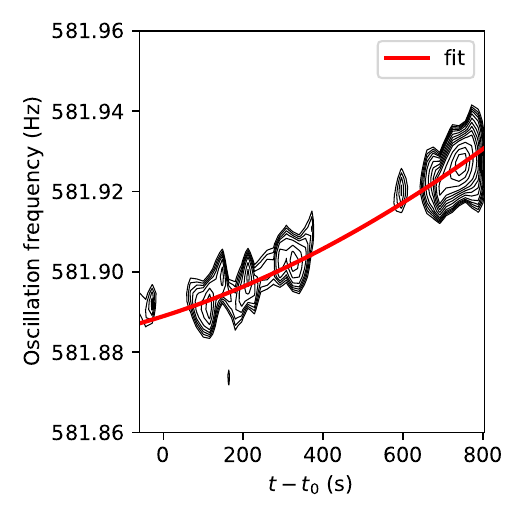}
      \caption{Dynamic $Z^2$ power spectrum for the data segment with pulsations. As in Figure ~3 in SM02, $Z^2$ statistics were computed within 64-s time intervals starting every 16 s. Contour grids correspond to $Z^2$ of 16, 18, 20, 22, 25, 30, 35, 40, 45, 50, 60, 70, 80, and 90. Red line shows $\nu(t)$ from Eq.~\ref{eq:nu_orb}. Time is measured from $t_0$ at 17:02:25 UTC on 2001 February 22. }
    \label{fig:Z2}
\end{figure}

To find the frequency of pulsations, we computed $Z^2$ statistics \citep{Buccheri:1983} in 64-s windows starting every 16 seconds. As in SM02, for $t_0$ we adopted the barycentric equivalent of topocentric timestamp UTC 2001-02-22 17:02:25. Contours for $Z^2$ are shown on Figure \ref{fig:Z2}. The oscillation frequency undergoes visible linear drift over the course of 800\,s. SM02 attributed this drift to the orbital motion of the NS and fitted it with the following expression:
\begin{equation}
\label{eq:nu_orb}
    \nu(t) = \nu_0\left[1-\frac{V_\mathrm{ns}\sin i}{c}\sin(2\pi (t-t_0)/P_\mathrm{orb} + 2\pi\phi_0)\right],
\end{equation}
where $\nu_0$ is the barycentric frequency, $V_\mathrm{ns}\sin i/c$
the projected neutron star velocity, $\phi_0$  the fractions of the orbital phase at the start of the pulsation interval, and $P_\mathrm{orb}$ the orbital period. 

Although our dataset is the same as in SM02 and our analysis follows the one in their work closely, we found that the drift is best described by a slightly different combination of parameters, which were obtained by maximizing $Z^2$ statistics over a grid of trial values. The largest difference was found to be in $\nu_0$, with our value of $\nu_0 = 582.1424$ \,Hz being 3\,mHz lower. The other parameters adopted were as follows: $V_\mathrm{proj} = 135.2$, $\phi_0 = 0.292$. All of these values lie within the $90\%$ confidence levels of the fit from SM02. Similarly to SM02, we fixed the value of the orbital period at $P_\mathrm{orb} = 3.79316732$\,hr obtained from optical observations.

Photon energies were calibrated with a response matrix created with the standard task \texttt{pcarsp} from the HEASARC package.
For energy-resolved analysis, we selected data from the channel subset [1, 30], corresponding to a nominal photon energy range of [2.0, 26.5]\,keV as this range contains the majority of the observed counts (see Figure \ref{fig:light_cures}). Background count rates were estimated from observations right before the burst start conducted in the same mode as the main data file and calibrated with the same response matrix.

\subsection{Model}\label{sec:model} 
The space-time model provides information on the mass and radius of the NS. In this work, we assume a Schwarzschild space-time while incorporating special relativistic effects and accounting for stellar oblateness due to rotation \citep[see,][for more detail]{Poutanen:2003, Morsink:2007,AlGendy:2014, Bogdanov:2019}. For the surface emission, we assume the coherent pulsations observed during the superburst were generated by a single, uniform-temperature hot spot. This model was identified by \citet{Artigue:2013} as a viable option for modelling the oscillations observed during the superburst of \name. We also assume a non-zero temperature for the rest of the stellar surface, as the entire surface is expected to undergo burning during the superburst. To track the changes in pulse fraction observed in the data, we divided the superburst data set into seven\footnote{The light curve was divided into seven segments to reflect changes in rms FA over time while keeping the computational cost manageable. The first 400 seconds (see Figure \ref{fig:Z2}), with strong oscillations, were split into four 100-second segments. The low-amplitude interval from about 400 to about 570 seconds was treated as a single segment, and the final portion, where oscillations re-emerge, was divided into two segments of approximately 150 seconds each.} different segments. We allowed the hot spot temperature and angular radius to evolve independently across these segments. However, because the cooling is relatively slow during the superburst (see Figure \ref{fig:light_cures}) and the pulsations occur only in a small portion of the cooling phase, we kept the temperature of the rest of the stellar surface non-variable across segments. Similarly, the phase shift was kept non-variable across segments, as the observed frequency drift arises from orbital motion rather than changes in the hot spot’s location. For computational efficiency, we also kept the co-latitude of the hot spot center non-variable across segments.

 We used the hot neutron star atmosphere model (solar composition) of \citet{Suleimanov:2012} to account for modifications to the spectrum and pulse shape as photons pass through the thin layer of atmosphere surrounding the neutron star. For the background model, we follow the standard procedure outlined in Section \ref{sec:intro}, using a range of \texttt{fa} values. Given the sensitivity of the results to the background limit, and thus the \texttt{fa} value, we considered six different scenarios. In \texttt{Case 0}, we set $1.5 \leq \texttt{fa} \leq 2$, consistent with the spectral modelling of \citet{Keek:2014}. For the remaining cases, we set a more conservative threshold by using the instrument background as the lower limit.  The upper limits were set as follows: \texttt{fa} = [1,2,3,4] for \texttt{Case [1,2,3,4]}. Values greater than 4 lead to negligible differences, as the pre-burst counts, scaled by \texttt{fa}, become nearly identical or exceed the data in most channels. Therefore, we group all cases with \texttt{fa} values above 4 into a single scenario: \texttt{Case 5}. In this case, no upper limit is imposed on the background, allowing the background count rate to reach the observed data. This is a natural and agnostic choice when one lacks prior information about the background contribution to the total counts \citep[see e.g.][]{Riley:2019}.

To account for absorption in the interstellar medium, we used the \texttt{Tbabs} model, which uses the photoelectric absorption cross-section from \citet{Wilms:2000}. For \texttt{Case [1,2,3,4,5]}, we kept the interstellar attenuation column density, $N_H$, non-variable across all segments. However, for \texttt{Case 0}, we allowed $N_H$ to vary between segments to facilitate comparison with the results of \citet{Keek:2014}.

For inference, we set the spin frequency of the neutron star to 581.91 Hz (for each segment), which corresponds to the average barycentric oscillation frequency (see SM02). We also use the instrument response extracted as described in Section \ref{sec:data}.

\subsection{Priors}\label{sec:prior} 
In Table \ref{tab:prior}, we provide a comprehensive list of all free parameters and their prior bounds used in this study. For the mass ($M$), equatorial radius ($R_{\rm eq}$), hot spot phase ($\phi_\mathrm{spot}$), hot spot co-latitude ($\Theta_\mathrm{spot}$), and hot spot angular radius ($\zeta_\mathrm{spotX}$, where X denotes the rank of the segment), we used the prior bounds as in previous PPM analyses \citep[see, e.g.,][]{Choudhury:2024, Kini:2024b,Salmi:2024, Vinciguerra:2024}. Detailed descriptions of the implicit priors affecting the joint ($M, R_{\rm eq}$) prior space can be found in Section 2.5 of \citet{Kini:2023} and Section 2.2.5 of \citet{Kini:2024b}. 

For the observer’s inclination ($\cos(i)$), we used the limits set by \citet{Casares:2006} obtained from Doppler tomography in the optical waveband. The bounds of hot spot temperature ($T_\mathrm{spotX}$) along with the star temperature ($T_\mathrm{starX}$) are set by the hot neutron star atmosphere model \citep[see Section 2.2 of ][for a detailed description]{Kini:2023}.

For the distance ($D$) prior, we used a uniform distribution ranging from 1.0 kpc to 20.0 kpc, despite existing distance prior knowledge about the optical companion star V801 Ara \citep{Gaia, Bailer-Jones:2021}. This is because the geometric distance and the photogeometric distance, although broadly consistent, differ significantly from each other (see Figure \ref{fig:prior} of Appendix \ref{sec:appendix}). The potential reasons for this difference are legion and beyond the scope of this work\footnote{Private discussion with Coryn Bailer-Jones. But some information can be found in \citet{Bailer-Jones:2021}  and at \url{https://www2.mpia-hd.mpg.de/~calj/gedr3_distances/FAQ.html}}. Given the computational constraints of using each distance prior for all of the cases, we adopted a more conservative approach by using a uniform distribution. The bounds of this distribution were defined by the lowest value of the geometric distance samples and the highest value of the photogeometric distance samples.

For the ISM column density ($N_{H}$), we used bounds consistent with \citet{Keek:2014} for \texttt{Case 0} to facilitate comparison. For the remaining cases, we used bounds consistent with the values from HI maps of \citeauthor{HI4PI_Collaboration} obtained with the \texttt{nH} tool\footnote{\url{https://heasarc.gsfc.nasa.gov/cgi-bin/Tools/w3nh/w3nh.pl}}.

\begin{table*}
    \centering
    \begin{threeparttable}
    \begin{tabular*}{\textwidth}{@{\extracolsep{\fill}} lll}
     \hline \hline
        Parameter & Description & Assumed prior\\
         \hline
          $M$ (\msol)  & Gravitational mass & $M\sim\mathcal{U}(1.0,3.0)$ \\  
          $R_{\rm eq}$ (km)  & Equatorial radius & $R_{\rm eq}\sim\mathcal{U}(3r_g(1.0),16.0)$\tnote{\textcolor{mygreen}{a}} \\
          $D$ (kpc)          & Distance & $D\sim\mathcal{U}(1.0,20.0)$ \\
          $\cos(i)$          & Cosine of the observer’s inclination & $\cos(i)\sim\mathcal{U}(\cos(74^\circ),\cos(30^\circ))$ \\
          $\phi_\mathrm{spot}$ (cycles)  & Hot spot phase & $\phi_\mathrm{spot}\sim\mathcal{U}(-0.25,0.75)$ \\
          $\Theta_\mathrm{spot}$ (radian) & Hot spot co-latitude & $\cos(\Theta_\mathrm{spot}) \sim\mathcal{U}(-1.0,1.0)$ \\
          $\zeta_\mathrm{spotX}$ (rad) & Hot spot angular radius & $\zeta_\mathrm{spot}\sim\mathcal{U}(0.0,\pi/2)$ \\
          $\log[T_\mathrm{spotX}(\mathrm{K})/1\mathrm{K}]$ & Hot spot temperature & $\log[T_\mathrm{spot}(\mathrm{K})/1\mathrm{K}]\sim\mathcal{U}(6.7,7.6)$\\
          $\log[T_\mathrm{starX}(\mathrm{K})/1\mathrm{K}]$ & Star temperature & $\log[T_\mathrm{star}(\mathrm{K})/1\mathrm{K}]\sim\mathcal{U}(6.7,7.6)$ \\
          
          \multirow{2}{*}{$N_{H}$ ($10^{20}\mathrm{cm}^{-2}$)} & \multirow{2}{*}{ISM column density} & $N_{H} \sim \mathcal{U}(15.0,500.0)$ for \texttt{Case 0}  \\
                  &                   & $N_{H} \sim \mathcal{N}(30.0,4.0^2)$ for \texttt{Case [1,2,3,4,5]} \\
          \hline
    \end{tabular*}
    \caption{List of all free parameters and their prior bounds used in this study.}
    \label{tab:prior}
    \begin{tablenotes}
      \item[a] $r_g(1.0)$: Solar Schwarzschild gravitational radius.
    \end{tablenotes}
    \end{threeparttable}
\end{table*}

\subsection{Posterior Computation}\label{sec:sampling_post}

To compute the posterior samples, we use \texttt{MultiNest} \citep{MultiNest_2008,MultiNest_2009,MultiNest_2019} and  \texttt{PyMultiNest} \citep{pymultinest:2014}. For all the cases, we set the sampling efficiency and evidence tolerance to 0.3 and 0.1, respectively \citep[for the definition of sampling efficiency and evidence tolerance, see][]{Vinciguerra:2023, Salmi:2024}. For computational efficiency, we disabled the multimode/mode-separation of \texttt{MultiNest}. To determine the optimal number of live points, we randomly selected \texttt{case 4} and performed inference with 8,000 and 16,000 live points. As expected, the joint mass-radius posteriors broadened with an increased number of live points. However, the changes are not too large, and the 1D posteriors remained almost unchanged (see Figure \ref{fig:LP} in Appendix \ref{sec:appendix}). This provided confidence to proceed with 8,000 live points for the remaining cases.

\FloatBarrier

\section{Results}\label{sec:result}

In this section, we primarily focus on the mass ($M$), radius ($R_{\rm eq}$), and compactness\footnote{$C=GM/R_{\rm eq}c^2$, where $G$ is the gravitational constant and $c$ is the speed of light.} ($C$), given their importance for dense matter studies. However, the inferred properties of the hot spot for each case are also shown in Appendix \ref{sec:appendix}. We also summarize the main results presented in the section in Table \ref{tab:summary}.

\begin{figure*}
    \centering
    \includegraphics[width=1.8\columnwidth]{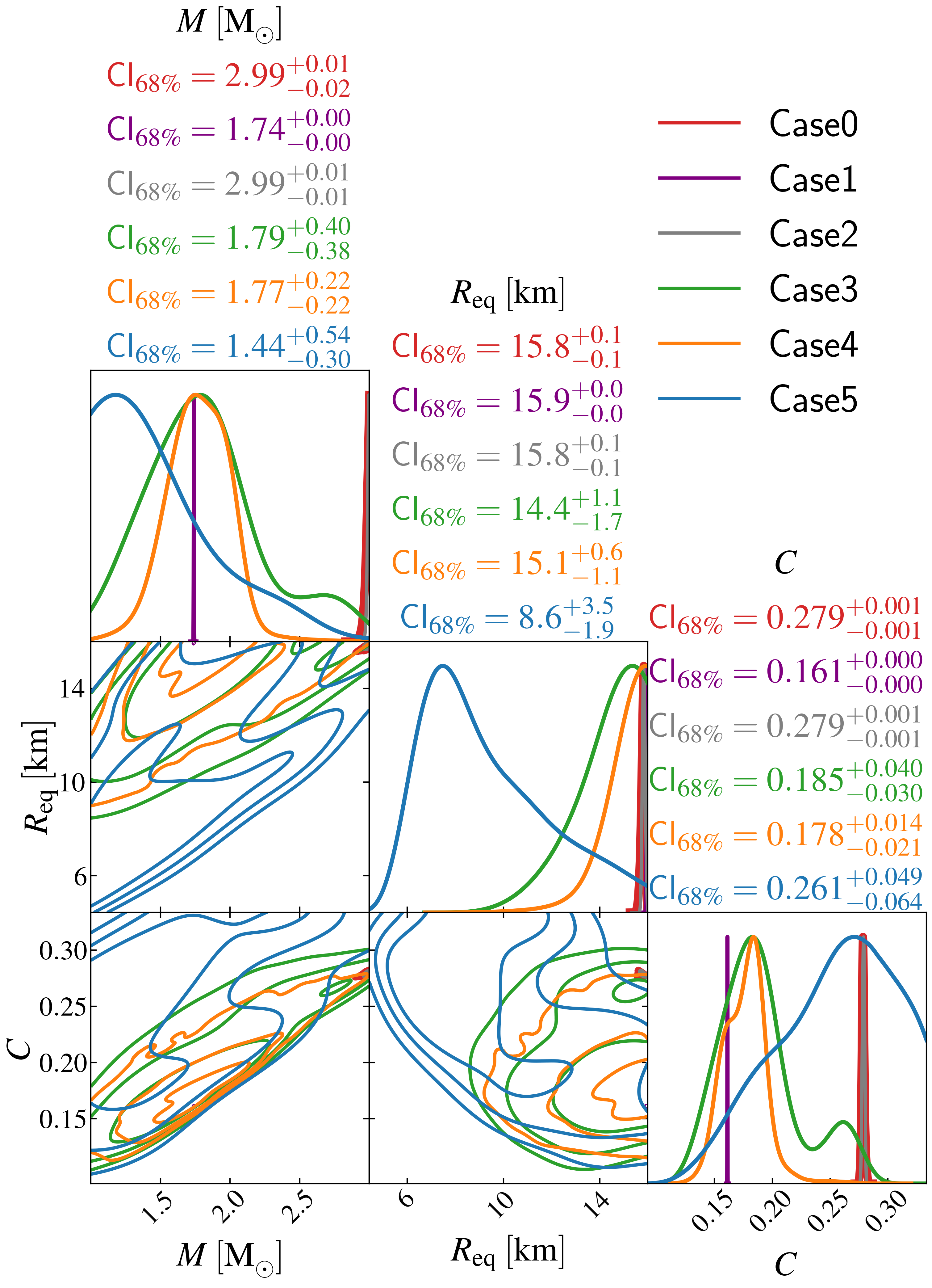}
    \caption{Posterior distributions of mass, radius, and compactness obtained under each background assumption. The 2D contours represent the $68\%$, $95\%$, and $99\%$ credible regions, while the 1D plots on the diagonals display the marginalized posterior distributions (unnormalized) of each parameter}
    \label{fig:mass_radius_compactness}
\end{figure*}

The posterior distributions of mass, radius, and compactness obtained under each background assumption are shown in Figure \ref{fig:mass_radius_compactness}. The 2D contours represent the $68\%$, $95\%$, and $99\%$ credible regions, while the 1D plots on the diagonals display the marginalized posterior distributions (unnormalized) of each parameter. Each background assumption leads to significantly different mass and radius configurations, ranging from a low mass and small radius in \texttt{Case 5} to a large mass and radius in \texttt{Cases 0,2}. Interestingly, \texttt{Case 3} and \texttt{Case 4} yield similar results. Likewise, \texttt{Case 0} and \texttt{Case 2} yield similar results, with both mass and radius in these cases approaching the upper bounds of the prior parameter space. The resulting compactness also varies widely, from low compactness in \texttt{Case 1} to very high compactness in \texttt{Cases 0,2,5}. 

The uncertainties in the mass and  radius for all cases except for  \texttt{Case 5} are particularly small. The uncertainties on the equatorial radius—$\Delta R_{\mathrm{eq}}/R_{\mathrm{eq}}$—are 0.5\%, 0.1\%, and  0.4\%, 10\%, and 5\% for \texttt{Case 0}, \texttt{Case 1}, \texttt{Case 2}, \texttt{Case 3} and \texttt{Case 4} respectively. Given that the total number of counts is approximately $8 \times 10^6$, with a background level of $\sim6\%$ (in the most favorable scenario, \texttt{Case 1}), and a root-mean-square (rms) fractional amplitude of only $\sim0.7\%$, such low uncertainties are implausible. Even under highly idealized conditions—such as an equatorial hotspot viewed edge-on, assuming a radius of 15\,km—the simplified estimation from \citet[][Eq.~4]{Psaltis:2014} yields a lower limit on the radius uncertainty of approximately 15\%. This discrepancy suggests that a significant portion of the parameter vectors within the prior space are unable to adequately describe the data under these background constraints \citep[see e.g.][]{Kini:2024a}. This may indicate that the background assumptions for these cases do not accurately reflect the actual conditions during the superburst. Moreover, the evidence for these cases is worse compared to the other cases (see  Table \ref{tab:evidence} of Appendix \ref{sec:appendix}), indicating that these background assumptions are not the preferred models.

\begin{figure*}
    \centering
    \includegraphics[width=0.47\textwidth]{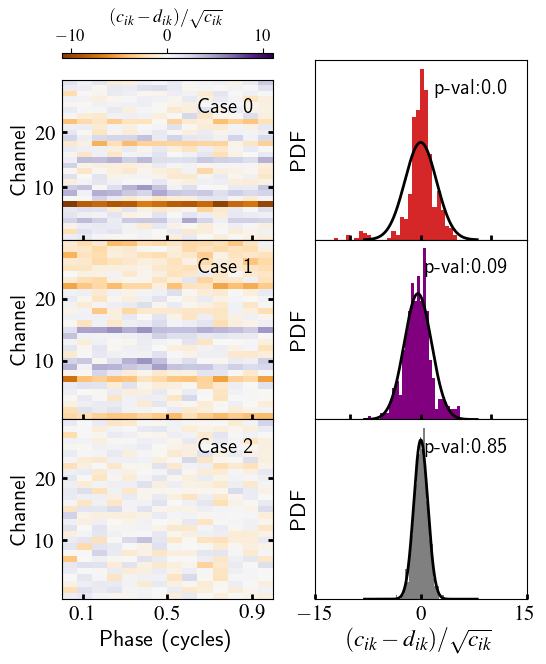}
    \hfill
    \includegraphics[width=0.47\textwidth]{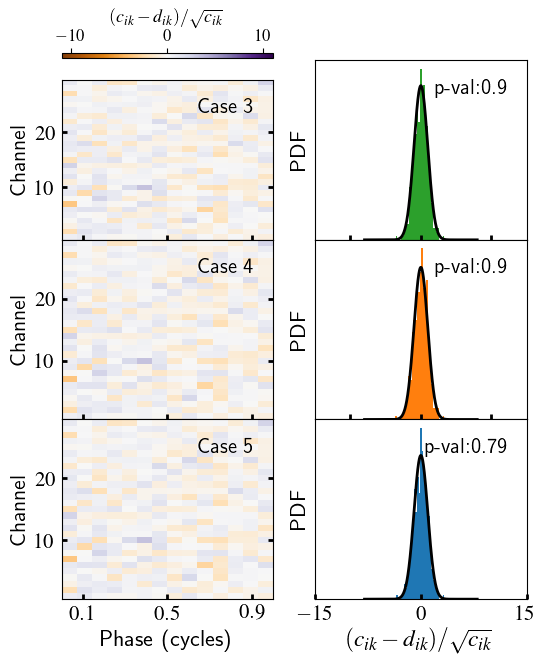}
    \caption{Residual plots and distributions for segment 4 for each case. The solid black line in the distribution plots indicates the normal distribution that best fits the residual distribution. Each residual distribution plot shows the p-value from the Kolmogorov-Smirnov (KS) test, assessing whether the residual distribution deviates from a normal distribution. $c_{ik}$ and $d_{ik}$ are respectively the model counts (signal averaged over 100 random samples from the posterior distribution) and the data counts in the $i^{\rm th}$ rotation phase and  $k^{ \rm th}$ energy channel.}
    \label{fig:residuals}
\end{figure*}

In Figure \ref{fig:residuals}, we show the residual plot for each case for segment 4 along with each residual distribution. Similar trends are observed in the other segments.  Each residual distribution figure includes the p-value (p-val) from the Kolmogorov-Smirnov (KS) test, which evaluates whether the distribution deviates from a normal distribution.

In \texttt{Case 0} and \texttt{Case 1}, distinct features appear in the residuals.  In some channels, the model predicts significantly fewer counts than are observed in the data, indicated by horizontal red bands. Conversely, in other channels, the model predicts more counts than are observed, shown in blue. The distinct features in the residuals across both cases show that the model, under these background constraints, struggles to accurately match the observed data, either underestimating or overestimating the counts. The KS test yields a p-value of  $3.28\times10^{-8}$ in \texttt{Case 0}, showing that the residual distribution deviates significantly from normality. This is not unexpected, given that the residual distribution is bimodal with the second mode caused by the model underestimating the observed counts in most channels. However, the p-value in \texttt{Case 1}
indicates that the residuals are marginally consistent with a normal distribution. In the remaining cases, however, the residuals do not exhibit such distinct features and their distributions are consistent with a Gaussian distribution (from the p-values).

\begin{figure}
    \centering
    \includegraphics[width=0.46\textwidth]{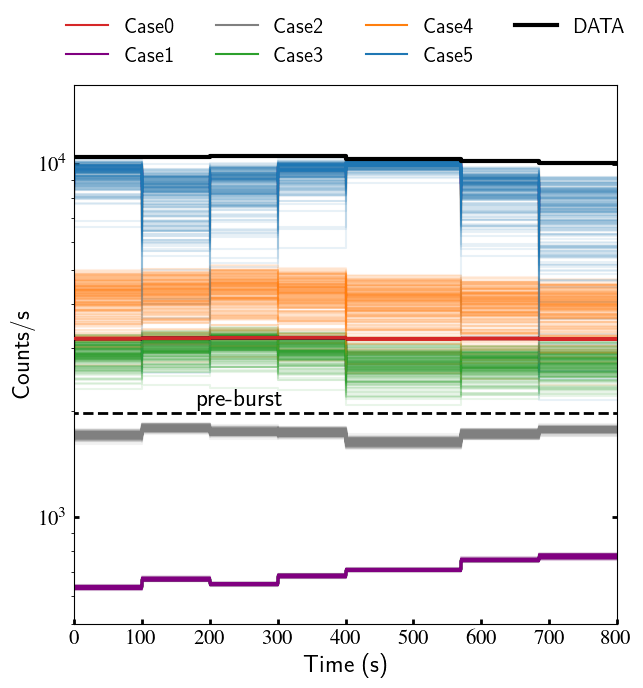}
    \caption{Inferred background count rate for 200 random samples, for each case alongside the observed data for each time segment. The "inferred background" is the background spectrum that maximizes the likelihood for a given parameter vector.   The horizontal black line indicates the pre-burst count rate}
    \label{fig:bkg}
\end{figure}

The inferred background count rate for 200 samples randomly selected from the posterior distribution is shown in Figure \ref{fig:bkg}. In both \texttt{Case 1} and \texttt{Case 2}, the inferred background remains below the pre-burst level (which is expected for \texttt{Case 1} and not necessarily for \texttt{Case 2}), indicated by the black dashed line. However, in \texttt{Case 5}, the inferred background is significantly higher, nearly matching the observed data in most segments. Specifically, for the maximum likelihood vector in \texttt{Case 5}, the inferred background count rate accounts for approximately 95\% of the actual data in segment 5. This indicates that, in this case, a significant portion of the data is inferred to be background.

\begin{table*}
    \centering
    \begin{threeparttable}
    \begin{tabular*}{\textwidth}{@{\extracolsep{\fill}} cccccc} 
     \hline \hline
        Case & Background limit            & $M$ (\msol) & $R_{\mathrm{eq}}$(km) & $C$ & Key results \\
         \hline
          0           & $1.5c^{\rm k}_{\rm pre-burst} + c^{\rm k}_{\rm instr}  \leq c^{\rm k}_{\rm model} \leq 2c^{\rm k}_{\rm pre-burst}$&  $2.99^{+0.01}_{-0.02}$  & $15.8^{+0.1}_{-0.1}$ & $0.279^{+0.001}_{-0.001}$ & 
          \makecell[l]{\textbullet~ Narrow posterior distributions \\ 
                       \textbullet~ Features in the residuals \\ 
                       \textbullet~ Residual distribution deviates from a normal distribution\\ 
                       } \\  \\  
          1           & $c^{\rm k}_{\rm instr}  \leq c^{\rm k}_{\rm model} \leq c^{\rm k}_{\rm pre-burst}$                    &  $1.74^{+0.00}_{-0.00}$  & $15.9^{+0.0}_{-0.0}$ & $0.161^{+0.000}_{-0.000}$ & 
          \makecell[l]{\textbullet~Narrow posterior distributions \\ 
                       \textbullet~Features in the residuals \\ 
                       \textbullet~Residual distribution is consistent with a normal distribution \\ 
                       } \\ \\ 
          2           & $c^{\rm k}_{\rm instr}  \leq c^{\rm k}_{\rm model} \leq 2c^{\rm k}_{\rm pre-burst}$                             &  $2.99^{+0.01}_{-0.01}$  & $15.8^{+0.1}_{-0.1}$ & $0.279^{+0.001}_{-0.001}$ & 
          \makecell[l]{\textbullet~No features in the residuals\\ 
                       \textbullet~Residual distribution is consistent with a normal distribution \\
                       \textbullet~Inferred background remains below pre-burst level}
                       \\ \\ 
          3           & $c^{\rm k}_{\rm instr}  \leq c^{\rm k}_{\rm model} \leq 3c^{\rm k}_{\rm pre-burst}$ &  $1.79^{+0.40}_{-0.38}$  & $14.4^{+1.1}_{-1.7}$ & $0.185^{+0.040}_{-0.030}$ & 
          \makecell[l]{\textbullet~No features in the residuals\\ 
                       \textbullet~Residual distribution is consistent with a normal distribution }
                       \\ \\ 
          4           & $c^{\rm k}_{\rm instr}  \leq c^{\rm k}_{\rm model} \leq 4c^{\rm k}_{\rm pre-burst}$                       & $1.77^{+0.22}_{-0.22}$   & $15.1^{+0.6}_{-1.1}$ & $0.178^{+0.014}_{-0.021}$ & 
          \makecell[l]{\textbullet~No features in the residuals\\ 
                       \textbullet~Residual distribution is consistent with a normal distribution} 
                       \\ \\
          5           & $c^{\rm k}_{\rm instr}  \leq c^{\rm k}_{\rm model} \leq c^{\rm k}_{\rm data}$                 &  $1.44^{+0.54}_{-0.30}$  & $8.6^{+3.5}_{-1.9}$  & $0.261^{+0.049}_{-0.064}$ & 
          \makecell[l]{\textbullet~No features in the residuals\\ 
                       \textbullet~Residual distribution is consistent with a normal distribution \\
                       \textbullet~Inferred background accounts for 95\% of observed counts \\ 
                       \quad in some segments} \\
    \hline
    \end{tabular*}
    \caption{Summary of background assumptions, along with the inferred mass, inferred radius, inferred compactness, and key observations for each case. $c^{\rm k}_{\rm pre-burst}$, $c^{\rm k}_{\rm instr}$, and $c^{\rm k}_{\rm model}$ represent the count rates in the $k^{\rm th}$ energy channel for the pre-burst, the instrumental background, and the model, respectively.}
    \label{tab:summary}
    \end{threeparttable}
\end{table*}

\FloatBarrier

\section{Discussion}\label{sec:discussion}

In this analysis, we used a single hot spot model along with various background assumptions—since the background levels and variation during a superburst are not known a priori—to infer the properties of \name. Our findings show that each background assumption resulted in distinct values for the inferred mass, radius, and compactness of \name. Similar results were obtained in the case of XTE~J1814$-$338 \citep{Kini:2024b}. These differences in compactness (hence in mass or radius) are not unexpected, as each background assumption inherently constrains the possible compactness of these sources. This is because for a given unpulsed component, setting a lower limit on the background implies that we allow the star to contribute more to the unpulsed component. This can be achieved by making the star more compact, ensuring that some part of the hot spot remains always visible to the observer due to gravitational lensing.

\subsection{Case 0}
 In \texttt{Case 0}, the distinct features in the residuals, the extremely narrow posterior distributions, and the deviation of the residual distribution from the expected Gaussian profile suggest that the models used (the single hot spot model along with the other models described in \ref{sec:model}) may not be sufficient to describe the data, assuming the background limits found in \citet{Keek:2014} accurately reflect the true background. This inadequacy mirrors findings for XTE~J1814$-$338 \cite[][]{Kini:2024b}. It indicates the probable need to incorporate additional components that can give rise to spectral lines, such as an accretion disk, into the model. During the superburst, intense X-rays emitted from the stellar surface illuminate the accretion disk, causing a significant portion of these X-rays to reflect off the disk. This reflection resulted in emission and absorption lines, along with reflection components, all of which have been identified in the data \citep{Keek:2014}. While adding these components might help explain the data at lower energy channels—particularly below channel 15, as the iron line appears around 6.5 keV and iron edges near 9 keV \citep{Keek:2014, Koljonen:2016}—they may not address the discrepancies observed in higher channels.  Additionally, the pronounced feature in channel 7 may not indicate an issue with that specific channel but rather a missing component in our model. This is because count excess and deficit can appear and disappear in different channels as more components are added, as found in the superburst spectral modelling of \name~ \citep[see Figure 1 in][]{Keek:2014}.

\subsection{Cases 1  and 2}

In \texttt{Case 1}, the features in the residuals and their distribution indicate poor model performance under this background assumption. While the residuals and p-values in \texttt{Case 2} suggest that the model performs reasonably well under this background assumption, the extremely small parameter space regions that can explain the data in both \texttt{Case 1} and \texttt{Case 2} suggest that these background assumptions might not accurately capture the true physics \citep[see][]{Kini:2024a}. Moreover, in \texttt{Case 2}, most samples yield an inferred background below the pre-burst level, even though the background was allowed to reach up to twice the pre-burst levels. This suggests that the likelihood hyper-volumes explored in both cases are similar, with each solution representing just one mode of the likelihood hyper-volume. The extremely small credible regions in both cases further indicate that the live points may have been stuck in local maxima during the runs \footnote{We also note that \texttt{Case 1} missed the better fitting \texttt{Case 2} solution (which is within the \texttt{Case 1} prior space). This implies that the live points were stuck for that case.}. Given this, the likelihood surface is likely multimodal, making it desirable to perform the runs with the multi-mode of MultiNest enabled, using more live points. However, due to current computational constraints, we were unable to explore runs with more live points or enable the multi-mode option in MultiNest.

In both cases, the inferred background remains below the pre-burst level, which could imply a decrease in the accretion rate during the superburst, at least over the time interval considered in this analysis. However, given the features in the residuals in \texttt{Case 1} and the extremely narrow posterior distributions in both cases, it is highly unlikely that the accretion rate decreased during the superburst.

\subsection{Cases 3  and 4}
The lack of clustering in the residuals and the fact that the residual distributions are consistent with a Gaussian distribution in \texttt{Case 3}, and \texttt{Case 4} suggest that setting a high upper limit on the background results in a better fit. This is because, in the background treatment in \texttt{X-PSI}, an excess or deficit of counts in a given channel in the model can be adjusted through background marginalization \citep[see Appendix B.2 of ][]{riley_thesis}, as long as this excess or deficit falls within the bounds set on the expected background from the source. During a burst, an increase in the mass accretion rate by up to a factor of 4 can be expected \citep{Fragile:2018, Fragile:2020, Speicher:2023, Speicher:2024}. While the radius uncertainties obtained in \texttt{Cases 3} and \texttt{4} are larger than those in \texttt{Case 0,1,2}, they still remain below the idealized lower limit estimated in Section~\ref{sec:result}. This suggests that, despite the improved fit, the results obtained under these background assumptions may still underestimate the true uncertainties and are therefore unlikely to represent the true physical solution.

\subsection{Case 5}

In \texttt{Case 5}, an excess or deficit of counts in a given channel in the model can be adjusted through background marginalization similar to \texttt{Case 3}, and \texttt{Case 4}. This explains the lack of clustering in the residuals. The inferred uncertainty on the radius is approximately 30\%, which is substantially larger than the idealized lower limit derived in Section~\ref{sec:result}. However, the inferred background is extremely high, reaching up to 95\% of the total observed counts in segment 5. This result is inconsistent with the current burst model. During a superburst, most photons are expected to originate from nuclear burning on the stellar surface rather than from a simple enhancement of accretion \citep{Zand:2017}.

\subsection{Caveats}\label{sec:caveat}
The background limits explored in this paper struggled to explain the data, in part due to several simplifying assumptions made throughout our analysis. One important caveat in our modelling is the assumption that the entire stellar surface is visible. However, since \name \ is not an accretion-powered pulsar, it is highly likely that the accretion disk is in contact with the stellar surface, potentially obscuring a significant portion of the star \citep[][]{Keek:2014b}.  In future analyses, we plan to incorporate the effects of the accretion disk on the observed X-ray spectrum, including emission and absorption lines, as well as reflection components.

Other simplifications may also affect our inferences. These include assuming a single hot spot with a fixed shape and colatitude across all segments, as well as a uniform temperature across the spot and stellar surface. These assumptions could lead to systematic differences in background inference and, consequently, in mass and radius estimates. Future extensions could explore multiple spots, non-circular spot shapes \citep[e.g., r-modes;][]{Heyl:2004, Lee:2004, Piro:2005,Chambers:2019, Chambers:2020}, and temperature gradients across the hotspot/stellar surface. However, these additions will significantly increase computational costs and are left for future work.

Another limitation arises from our choice of segmenting the superburst light curve into seven time intervals. While this scheme was designed to capture the observed variation in pulse amplitude (see Figures ~\ref{fig:light_cures} and ~\ref{fig:Z2}) and to keep the parameter space tractable, it remains possible that the segmentation itself introduces biases. We did not systematically explore alternative segmentation strategies due to computational constraints, and thus the sensitivity of our results to the chosen scheme remains an open question.

Finally, a further limitation comes from disabling \texttt{MultiNest} multi-mode functionality, which may have resulted in live points becoming trapped in local maxima. Additionally, the parameter space exploration might not have been thorough enough, leading to different solutions for different cases—a challenge similar to that encountered in \citet{Salmi:2024_submitted} during PPM on PSR~J1231$-$1411. While increasing the number of live points and reducing \texttt{MultiNest}'s sampling efficiency could potentially mitigate this issue, it comes with significantly increased computational costs. Given that the current runs already require substantial computation time (about $1.52 \times 10^6$ core hours for all the runs), further optimization presents a considerable challenge.

Although incorporating additional physical effects—such as non-circular spot shapes, disk occultation, reflection, etc—would increase model complexity and the number of free parameters, we argue that the primary limitation is not the data quality. The dataset contains approximately $8\times10^6$ counts, which is significantly more than those used in previous PPM analyses of rotation-powered pulsars \citep[see e.g.][]{Choudhury:2024,Dittmann:2024,Salmi:2024, Vinciguerra:2024}. Even if only a small fraction of this emission originates from the hot spot, this still represents a sufficient number of counts to allow for meaningful constraints. Instead, we find that the dominant limitations arise from astrophysical uncertainties, particularly in the background modeling, and from the computational cost of exploring high-dimensional parameter spaces.

\section{Conclusion}\label{sec:conclusion}

In this study, we applied PPM to the pulsations observed during the 2001 superburst of \name, exploring various background assumptions to infer the mass, radius, and compactness of the neutron star. Our findings show that the inferred properties are highly sensitive to the background assumptions, resulting in significant variations across different scenarios. This underscores the critical need for a more accurate understanding of the background contributions during thermonuclear bursts to improve the reliability of PPM analyses of burst oscillations. 

For \texttt{Cases 0} and \texttt{1}, features in the residuals and the small credible regions suggest that the corresponding background assumptions do not accurately reflect the physical conditions during the superburst. \texttt{Cases 2}, \texttt{3}, \texttt{4}, and \texttt{5} yield better fits, as indicated by improved residuals and p-values. However, in \texttt{Cases 2}, \texttt{3}, and \texttt{4}, the credible intervals remain significantly narrower than expected, indicating that these background assumptions also fail to fully capture the underlying physics. In \texttt{Case 5}, although the fit also improves, the inferred background level is extremely high—reaching up to 95\% of the total observed counts—which is inconsistent with the current understanding of superburst emission, where most of the observed flux is expected to originate from nuclear burning on the stellar surface.

The challenges in explaining the data under the background assumptions explored in this paper, may stem from several assumptions discussed in Section \ref{sec:caveat}. Looking ahead, we aim to adopt more sophisticated models for the accretion disk and its interactions with the neutron star during superbursts. A key challenge will be gaining a clearer understanding of the accretion contribution during thermonuclear bursts, which is essential for accurately inferring the properties of burst oscillation sources using pulse profile modelling. Another promising direction would be to constrain the mass and radius prior space by using an equation of state-informed prior, similar to the approach in \citet{Salmi:2024_submitted}. Furthermore, obtaining independent and reliable constraints from burst spectra modelling \citep[e.g.][]{Suleimanov:2011, Suleimanov:2012, Ozel:2013, Steiner:2013, Guillot:2014, Nattila:2016, Nattila:2017, Suleimanov:2017a, Steiner:2018, Molkov:2024} or burst property modelling \citep[e.g.][]{Goodwin:2021, Galloway:2024} could greatly aid in constraining either the mass and radius or the superburst and the burst oscillation properties. Additionally, future work may benefit from leveraging more efficient inference techniques or machine learning-based surrogate models \citep[see e.g.][]{Miller:2022shs, sbi} to reduce computational costs. These developments could eventually make it feasible to explore more complex surface patterns and jointly model multiple bursts or superbursts from 4U 1636–536, under the assumption of shared intrinsic parameters such as mass, radius, distance, etc. Such a joint approach could, in principle, yield tighter constraints by mitigating source-intrinsic degeneracies and background-related uncertainties that affect individual burst analyses.

Traditionally, the accretion contribution during bursts is modelled by scaling the persistent emission by a factor, \texttt{fa}. In the hard state, the increased X-ray burst emission alters the spectral properties of the accretion disk \citep{Kajava:2017}, suggesting that \texttt{fa} may be energy-dependent—a factor we did not account for in this analysis. Additionally, a recent study \citet{Speicher:2024} indicates that \texttt{fa} correlates more strongly with the thermal temperature of the disk surface than with the mass accretion rate, complicating efforts to constrain the background. Identifying reliable tracers of the mass accretion rate through numerical simulations will be crucial for improving these constraints. This, in turn, will enhance our ability to probe the properties of dense matter and may provide new insights into the origin of burst oscillations.

\section*{Acknowledgements}
We are grateful to the anonymous referee for their thoughtful comments and constructive suggestions, which helped improve the clarity and depth of the manuscript.
We also thank Coryn Bailer-Jones for the discussion on distance priors and Tod E. Strohmayer for insightful discussions and feedback on the manuscript. YK, ALW, TS, and SV acknowledge support from ERC Consolidator Grant No.~865768 AEONS (PI Watts). D.R.B. acknowledges support from NASA award 80NSSC24K0212 and NSF grant AST-2407658.

\section*{Data availability}
All data, including posterior samples, as well as the scripts used for the analyses, runs, and plots, are publicly available on Zenodo(\url{https://doi.org/10.5281/zenodo.14850910}).

\bibliographystyle{aasjournal}
\bibliography{bibliography}

\begin{appendix}
\section{Priors, Posteriors, and Evidence Analysis}\label{sec:appendix}

\begin{figure}
	\includegraphics[width=1\columnwidth]{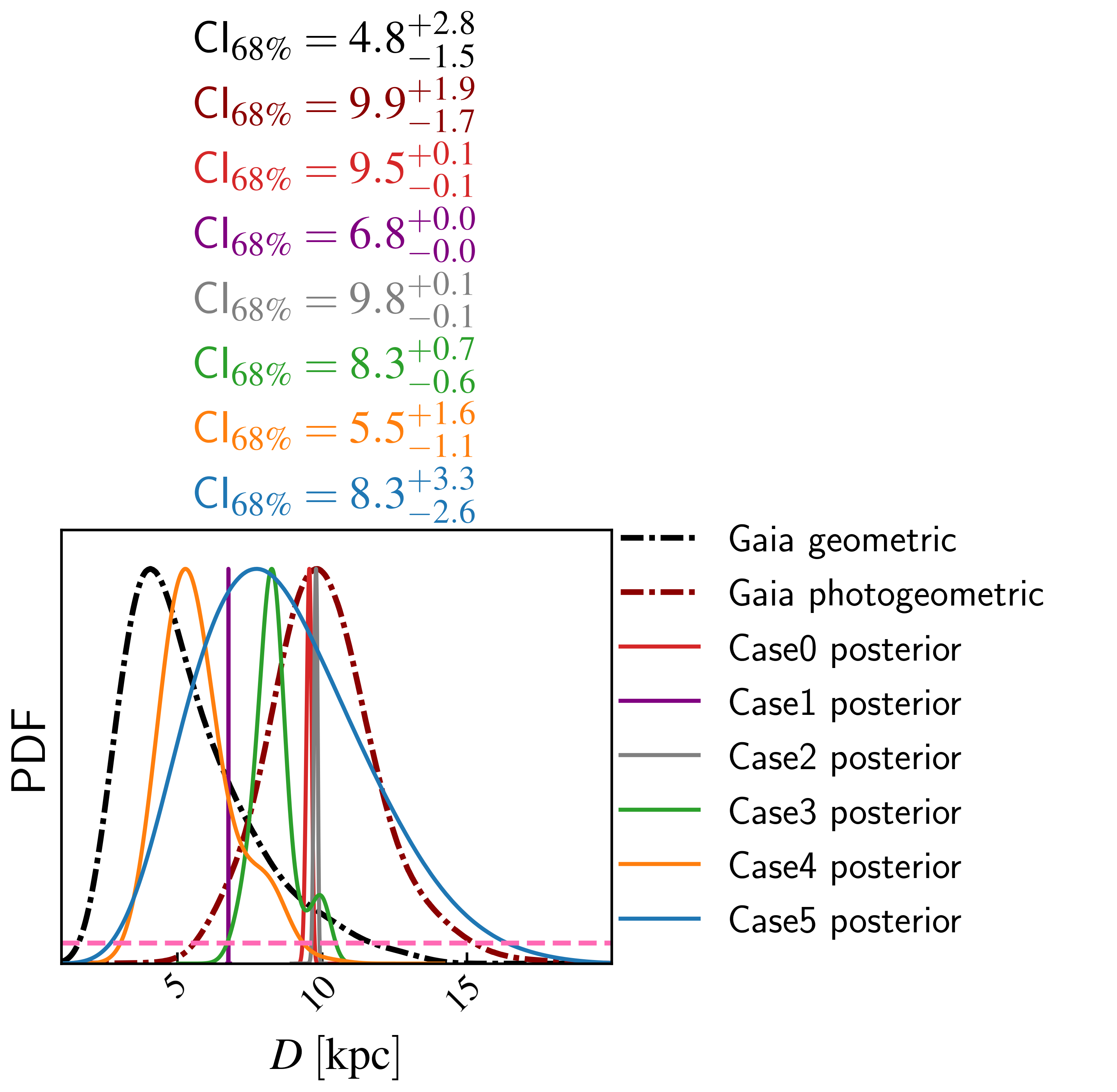}
    \caption{Comparison of the Gaia geometric and photogeometric distances (not used in this work) with the inferred posterior distribution for each case. The uniform prior used in this study is shown by the hot pink dashed line.}
    \label{fig:prior}
\end{figure}

Figure \ref{fig:prior} shows the geometric and photogeometric distance distribution derived from Gaia DR3 \citep{Gaia, Bailer-Jones:2021} (not used in this work), alongside the uniform prior used in this work (hot-pink dashed line) and the inferred posterior distributions. Note that the distributions are not normalized\footnote{The area under the curve is not necessarily equal to 1.}.  Although the geometric and photogeometric distance distributions differ significantly, they remain broadly consistent. In \texttt{Cases 0} and  \texttt{Cases 2}, the inferred posterior distributions are only marginally consistent with the Gaia geometric distance distribution, intersecting only at the tail of the distribution. In  \texttt{Case 1}, the inferred distribution is marginally consistent with the Gaia photogeometric distance distribution. However, in the remaining cases, the inferred distributions are consistent with both the geometric and photogeometric distance distributions.
 
\begin{figure}
	\includegraphics[width=1\columnwidth]{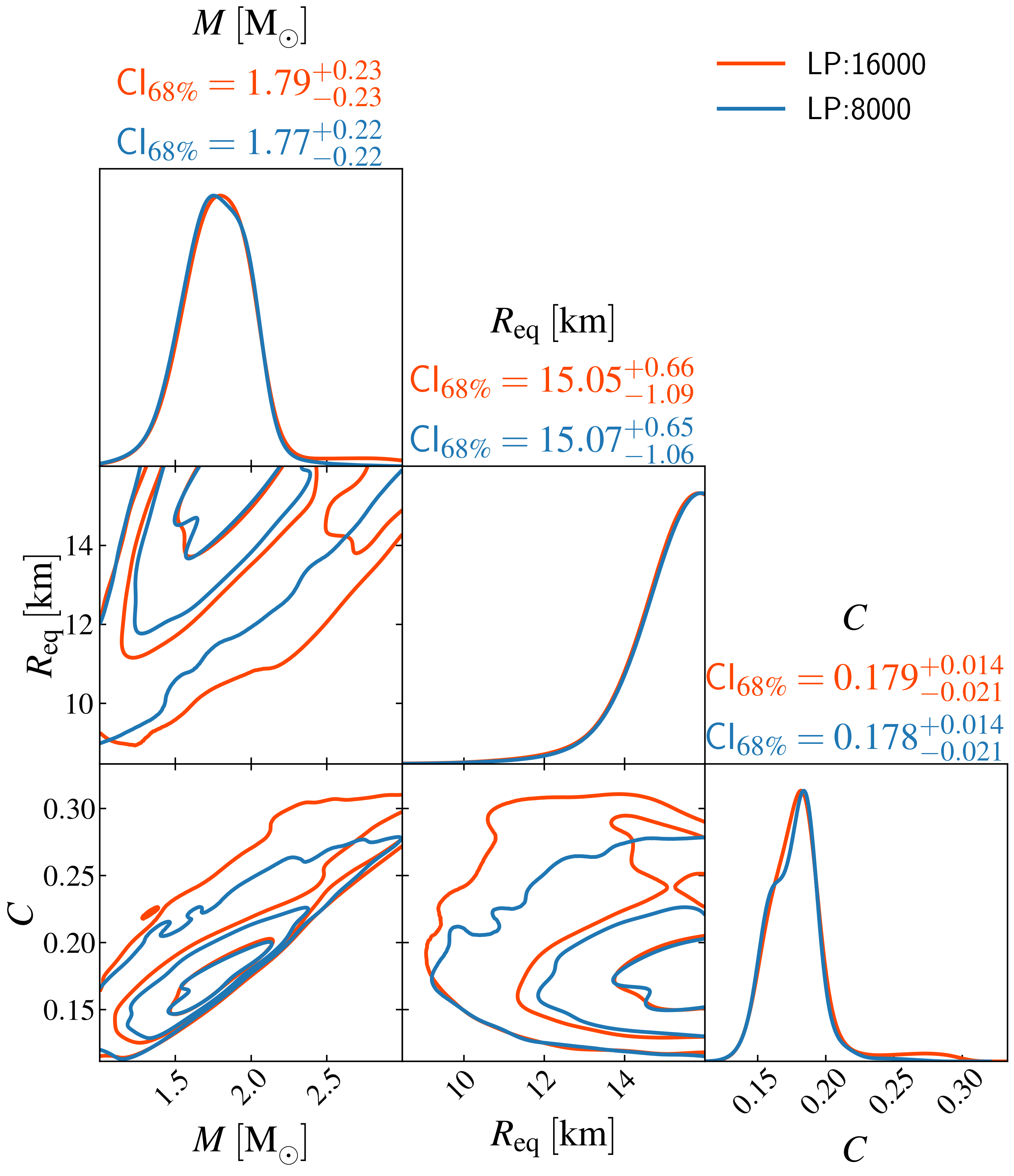}
    \caption{Posterior distributions of mass, radius, and compactness computed using varying numbers of live points (\texttt{Case 4)}.}
    \label{fig:LP}
\end{figure}

In Figure \ref{fig:LP}, we show the posterior distributions of mass, radius, and compactness computed with varying numbers of live points for \texttt{Case 4}. The posterior distributions for the remaining parameters are available on Zenodo (\url{https://doi.org/10.5281/zenodo.14850910}). The joint mass-radius posteriors became broader with more live points. However, the 1D posteriors remained largely unaffected. A similar trend can be seen (see Zenodo) for the remaining parameters, which reinforced our confidence in using 8,000 live points for the remaining cases.

We report in Table \ref{tab:evidence} the Nested Sampling global ln-Evidence for each background assumption. Based on the ln-Evidence, \texttt{Case 4} emerges as the best model. However, the difference in ln-Evidence between \texttt{Case 4} and \texttt{Case 3} or \texttt{Case 5} is minimal (only 10 in natural logarithmic units). This indicates that the models are almost indistinguishable in their ability to explain the data.

\begin{table}
    \centering
    \begin{tabular}{cc}
    \hline \hline
        Case            &  ln-Evidence ($\times 10^5)$\\ 
    \hline
       \texttt{Case 0}  & -0.2108\\
       \texttt{Case 1}  & -0.1923 \\
       \texttt{Case 2}  & -0.1435 \\
       \texttt{Case 3}  & -0.1422 \\
       \texttt{Case 4}  & -0.1421 \\
       \texttt{Case 5}  & -0.1422  \\
       \hline 
    \end{tabular}
    \caption{Nested Sampling global ln-Evidence for each background assumption.}
    \label{tab:evidence}
\end{table}

\begin{figure}
	\includegraphics[width=1\columnwidth]{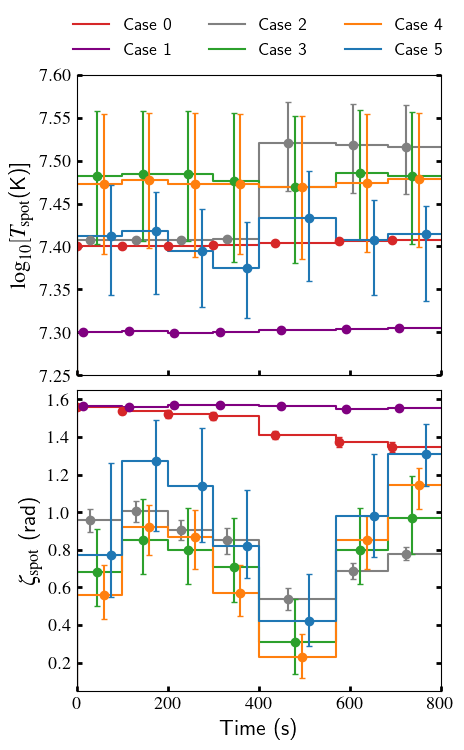}
    \caption{Median of the inferred temperature and angular radius of the hot spot. The error bars are the 68\% uncertainty regions.}
    \label{fig:profiles}
\end{figure}

Figure \ref{fig:profiles} shows the inferred evolution of the temperature and angular radius of the hot spot during the superburst. In most cases (except \texttt{Case 2}), the inferred temperature of the hot spot remains generally constant throughout the superburst.  In \texttt{Case 1}, the angular radius of the hot spot also stays constant during the superburst. Therefore it is not surprising why \texttt{Case 1} struggles to explain the data given that the observed rms FA varies during the burst.

For the other cases, except \texttt{Cases 0,1}, a general trend is observed in the hot spot's angular radius. It stays relatively stable (within the 68\% level) during the initial part of the burst, then decreases before increasing again over time. This trend aligns with the observed rms FA and implies that the variation in the rms FA is due to changes in the hot spot's angular radius.

If \name \ is indeed accreting via a boundary layer, then changes in the hot spot size might be attributed to the accretion disk wobbling during the superburst. This wobble could make parts of the hot spot more or less visible over time. However, it remains uncertain whether a disk could wobble over such a short time scale. Another possible explanation is that the disk itself pulsates (e.g. the inner disk moving in and out) during the superburst, though this seems unlikely since the burst was not a PRE burst \citep{Keek:2014}. Alternatively, the inner accretion disk, depleted at the start of the burst, could be replenished, making the inner disk more opaque and thus causing the hot spot to become less visible at certain points during the burst, and more visible again if the inner disk depletes once more. However, replenishment would imply an increase in the accretion rate, which should correspond to a rise in the inferred background during that period. However, this scenario is unlikely as the inferred background generally remains constant for most cases.

\end{appendix}

\end{document}